\documentstyle[11pt,epsf,fleqn]{article}

\begin{document}

\title{\bf{Generation of single-mode SU(1,1) intelligent states
and an analytic approach to their quantum statistical properties}} 
\author{{\Large
C Brif\thanks{Electronic address: costya@physics.technion.ac.il} \
and A Mann\thanks{Electronic address: ady@physics.technion.ac.il} } 
\vspace*{3mm} \\
{Department of Physics, Technion -- Israel Institute of 
Technology, Haifa 32000, Israel}}
\date{}
        \maketitle



        \begin{abstract}
We discuss a scheme for generation of single-mode photon states
associated with the two-photon realization of the SU(1,1) algebra.
This scheme is based on the process of non-degenerate down-conversion 
with the signal prepared initially in the squeezed vacuum state and 
with a measurement of the photon number in one of the output modes. 
We focus on the generation and properties of single-mode SU(1,1) 
intelligent states which minimize the uncertainty relations for 
Hermitian generators of the group. Properties of the intelligent 
states are studied by using a ``weak'' extension of the analytic 
representation in the unit disk. Then we are able to obtain exact 
analytical expressions for expectation values describing quantum 
statistical properties of the SU(1,1) intelligent states.
Attention is mainly devoted to the study of photon statistics
and linear and quadratic squeezing.
        \end{abstract}

\vspace{0.1cm}
\noindent \hspace{0.8cm}
{PACS numbers: 42.50.Dv, 03.65.Fd}


\section{Introduction}
\label{sec:intro}
\setcounter{equation}{0}

Intelligent states are quantum states which minimize uncertainty 
relations for non-commuting quantum observables [1--3].
In the last years there exists a great interest in various properties, 
applications and generalizations of intelligent states [4--22].
One of the reasons for this interest is the close relationship between 
intelligent states and squeezing. For example, the generalized 
intelligent states of the Weyl-Heisenberg group coincide with the
canonical squeezed states. In fact, the generalized intelligent states 
for two quantum observables can provide an arbitrarily strong squeezing 
in either of them \cite{Trif}. Therefore, a generalization of
squeezed states for an arbitrary dynamical symmetry group leads to the
intelligent states for the group generators \cite{NiTr,Trif}. In 
particular, the concept of squeezing can be naturally extended to the 
intelligent states associated with the SU(2) and SU(1,1) Lie groups.  
An important possible application of squeezing properties of the SU(2) 
and SU(1,1) intelligent states is the reduction of the quantum noise in 
spectroscopy \cite{AgPu94} and interferometry \cite{HM93,Brif:interf}.

In the present paper we propose a scheme for generation of intelligent 
photon states associated with the two-photon realization of the SU(1,1)
algebra. This scheme is based on a combination of degenerate and
non-degenerate optical parametric processes, with a measurement of the 
photon number in one of the output modes. In order to study properties 
of the intelligent states, we use a ``weak'' extension of the analytic 
representation in the unit disk \cite{BVM96}. In the context of the 
two-photon realization, this analytic representation is based on the 
single-mode squeezed states. The bases of the squeezed
vacuum states and the squeezed ``one photon'' states are used in the 
even and odd sectors of the Fock space, respectively. Thus we obtain 
the analytic representations of various eigenstates associated with 
the SU(1,1) algebra. Using these representations, we derive
exact closed expressions for moments of the SU(1,1) generators.
In this way we are able to study various quantum statistical properties
of photon states whose generation is possible in the presented scheme. 
We devote attention to the examination of photon statistics
and linear and quadratic squeezing. We show that the single-mode 
SU(1,1) intelligent states exhibit interesting nonclassical properties 
such as sub-Poissonian photon statistics and strong squeezing.

\section{The SU(1,1) intelligent states}
\label{sec:defin}
\setcounter{equation}{0}

In this work we consider two-photon parametric processes. Hamiltonians 
describing these processes involve operators of the form $a^2$ and 
$a^{\dagger 2}$, where $a$ is the annihilation operator of a quantized 
light mode. It is convenient to use the operators
\begin{equation}
K_{+} = \mbox{$\frac{1}{2}$} a^{\dagger 2} \hspace{0.8cm}
K_{-} = \mbox{$\frac{1}{2}$} a^2 \hspace{0.8cm}
K_{3} = \mbox{$\frac{1}{2}$} a^{\dagger} a + \mbox{$\frac{1}{4}$} 
\label{realiz}
\end{equation}
which form the single-mode two-photon realization of the SU(1,1) Lie
algebra,
\begin{equation}
[ K_{-} , K_{+} ] = 2K_{3} \hspace{0.8cm}
[ K_{3} , K_{\pm} ] = \pm K_{\pm} \, .
\end{equation}
One can also use the Hermitian combinations
$K_{1} = \frac{1}{2} (K_{+} + K_{-} )$ and 
$K_{2} = \frac{1}{2 {\rm i} } (K_{+} - K_{-} )$
which satisfy the well-known SU(1,1) commutation relations.
The Casimir operator $K^2$ for any irreducible
representation is the identity operator multiplied by a number,
\begin{equation}
K^2 = K_{3}^2 - K_{1}^2 - K_{2}^2 = k(k-1) I \, .   \label{casimir}
\end{equation}
Thus a representation of SU(1,1) is determined by the number $k$ called 
the Bargmann index \cite{Barg47}. The representation space ${\cal H}_{k}$ 
is spanned by the orthonormal basis $|n,k\rangle$ ($n=0,1,2,\ldots$). For 
the two-photon realization (\ref{realiz}), one obtains $K^{2} = -3/16$. 
Therefore, there are two irreducible representations: $k=1/4$ and $k=3/4$ 
\cite{BiVo87}. The representation space ${\cal H}_{e}$ ($k=1/4$) is the 
even Fock subspace with the orthonormal basis consisting of even number 
states $|n,\frac{1}{4}\rangle = |2n\rangle$ ($n=0,1,2,\ldots$); the 
representation space ${\cal H}_{o}$ ($k=3/4$) is the odd Fock subspace 
with the orthonormal basis consisting of odd number states 
$|n,\frac{3}{4}\rangle = |2n+1\rangle$ ($n=0,1,2,\ldots$).

Any two quantum observables (Hermitian operators in the Hilbert space) 
$A$ and $B$ obey the generalized uncertainty relation 
\begin{equation}
(\Delta A)^{2} (\Delta B)^{2} \geq \frac{1}{4} 
\left( \langle C \rangle^{2} + 4 \Delta_{AB}^{2} \right) 
\hspace{0.8cm} C = - {\rm i} [A,B]    \label{SR-unrel}
\end{equation}
where the variance of $A$ is $(\Delta A)^{2} = \langle A^{2} \rangle
- \langle A \rangle^{2}$, $(\Delta B)^{2}$ is defined similarly, the 
covariance of $A$ and $B$ is $\Delta_{AB} = \frac{1}{2} 
\langle AB+BA \rangle - \langle A \rangle \langle B \rangle$, and the 
expectation values are taken over an arbitrary state in the Hilbert 
space. When the covariance of $A$ and $B$ vanishes, $\Delta_{AB} = 0$, 
the generalized uncertainty relation (\ref{SR-unrel})
reduces to the ordinary uncertainty relation
\begin{equation}
(\Delta A)^{2} (\Delta B)^{2} \geq \frac{1}{4} 
\langle C \rangle^{2} \, .   \label{HR-unrel}
\end{equation}
Ordinary and generalized intelligent states provide an equality in the 
ordinary and generalized uncertainty relations (\ref{HR-unrel}) and
(\ref{SR-unrel}), respectively \cite{Trif}. 
The intelligent states for operators $A$ and $B$ are 
determined by the eigenvalue equation \cite{Trif,Puri}
\begin{equation}
(\eta A + {\rm i} B) |\psi\rangle = \lambda |\psi\rangle \ 
\label{is-def}
\end{equation}
where $\lambda$ is a complex eigenvalue. The parameter $\eta$ is 
complex for the generalized intelligent states and real for the 
ordinary ones. For ${\rm Re}\, \eta \neq 0$, expectation values over 
the intelligent states satisfy \cite{Trif}:
\begin{equation}
\langle A \rangle = \frac{ {\rm Re}\, \lambda }{{\rm Re}\, \eta } 
\hspace{0.8cm} 
\langle B \rangle = {\rm Im}\, \lambda - 
\frac{ {\rm Im}\, \eta }{ {\rm Re}\, \eta }\, {\rm Re}\, \lambda 
\label{intprop1}
\end{equation}
\begin{equation}
(\Delta A)^{2} = \frac{\langle C \rangle}{2{\rm Re}\, \eta} 
\hspace{0.8cm} 
(\Delta B)^{2} = \frac{|\eta|^{2} \langle C \rangle}{
2{\rm Re}\, \eta}   \hspace{0.8cm} 
\Delta_{AB} = - \frac{{\rm Im}\, \eta}{2{\rm Re}\, \eta} 
\langle C \rangle \, .   \label{intprop2}
\end{equation}

In the present work we consider the generation of the single-mode
SU(1,1) intelligent states. According to the above definitions,
the eigenvalue equation 
$(\eta K_{2} + {\rm i} K_{3}) |\psi\rangle = \lambda |\psi\rangle$
with real $\eta$ determines the ordinary intelligent states for the 
SU(1,1) generators $K_{2}$ and $K_{3}$. It means that these states 
provide an equality in the uncertainty relation 
$( \Delta K_{2} )^2 ( \Delta K_{3} )^2 \geq \mbox{$\frac{1}{4}$} 
\langle K_{1} \rangle^2$.
Analogously, the eigenvalue equation 
$(\eta K_{1} - {\rm i} K_{2}) |\psi\rangle = \lambda |\psi\rangle$
with real $\eta$ determines the ordinary intelligent states for the 
SU(1,1) generators $K_{1}$ and $K_{2}$. These states provide an 
equality in the uncertainty relation 
$( \Delta K_{1} )^2 ( \Delta K_{2} )^2 \geq \mbox{$\frac{1}{4}$} 
\langle K_{3} \rangle^2$.

\section{The generation scheme}
\label{sec:gener}
\setcounter{equation}{0}

Some schemes for the experimental production of the SU(2) and SU(1,1) 
intelligent states in nonlinear optical processes have been suggested 
recently \cite{PrAg,GeGr,LP96}. We focus on the most recent scheme, 
developed by Luis and Pe\v{r}ina \cite{LP96}, which employs two
important quantum mechanical features. The first one is the entangled
nature of the two-mode field generated by parametric down-conversion, 
and the second one is the role of measurement as a way to manipulate
the state of an entangled quantum system. 

It is known that in non-degenerate parametric down-conversion a 
measurement in the idler mode can be used to affect the state of the 
signal mode [26--31].
In particular, near-number states can be obtained in the signal mode 
by the measurement of the photon number in the idler mode \cite{HoMa86}. 
Luis and Pe\v{r}ina \cite{LP96} consider in their scheme two parametric 
down-conversion crystals with aligned idler beams and show that the 
measurement of the photon number in some of the modes leads to states 
which are related to the two-mode SU(2) and SU(1,1) coherent and 
intelligent states. The basic idea of this method is related to an 
interference experiment with signal beams coming from two parametric 
down-conversion crystals with aligned idler beams \cite{Mand,KSC94}.

In the present paper we consider a modification of the Luis-Pe\v{r}ina
scheme, which is suitable for the generation of the single-mode SU(1,1)
intelligent states. To this end, we examine the states produced in
parametric down-conversion with the signal mode prepared in the squeezed 
vacuum state, after the measurement of the photon number in one of the 
output modes. We show that this simple scheme produces the eigenstates
of a linear combination of the two-photon SU(1,1) generators. 
An additional SU(1,1) transformation, implemented by a degenerate 
parametric amplifier, takes these states into the single-mode SU(1,1) 
intelligent states.

The scheme under discussion is outlined in figure~1.
Two light beams represented by the mode annihilation operators $a$ and 
$b$ are mixed in the non-degenerate parametric amplifier NPA. 
The mode $a$ is beforehand squeezed in the degenerate parametric 
amplifier DPA. We assume that both parametric amplifiers are coherently
pumped by strong and undepleted classical fields, and that the system
is free of losses.

\begin{figure}[htbp]
\begin{minipage}[c]{.46\linewidth}
\epsfxsize=0.80\textwidth
\centerline{\epsffile{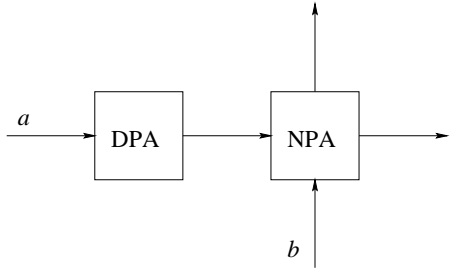}}\vspace*{0.2cm}
\end{minipage}
\begin{minipage}[c]{.46\linewidth}
\caption{Outline of the generation scheme. The light mode
$a$ is squeezed in the degenerate parametric amplifier DPA. Then the 
two light modes $a$ and $b$ are mixed in the non-degenerate parametric 
amplifier NPA.}
\label{fig:scheme}
\end{minipage}
\end{figure}

The processes of degenerate and non-degenerate parametric amplification 
are described by effective interaction Hamiltonians $H_1$ and $H_2$,
respectively,
\begin{equation}
H_1 = \mbox{$\frac{1}{2}$} g_1 a^{\dagger 2} + \mbox{$\frac{1}{2}$}
g_{1}^{\ast} a^2 \hspace{0.8cm} 
H_2 = g_2 a^{\dagger} b^{\dagger} + g_{2}^{\ast} a b 
\end{equation}
where $\hbar = 1$ and $g_1$, $g_2$ are parameters depending on the pump 
and the nonlinear characteristics of the media. The output state 
$|\psi\rangle$ of the whole system is related to the input state 
$|\psi\rangle_{\rm in}$ by the unitary transformation $U$,
\begin{equation}
|\psi\rangle = U |\psi\rangle_{\rm in} \hspace{0.8cm} 
U = U_2 U_1 = \exp(- {\rm i} H_2 t_2) \exp(- {\rm i} H_1 t_1) 
\end{equation}
where $t_1$, $t_2$ are the interaction times. We consider the input field 
in the vacuum state, $|\psi\rangle_{\rm in} = |0\rangle_a |0\rangle_b$. 
The mode $a$ is squeezed in DPA, and it enters NPA in the squeezed vacuum 
state. There is no need to take into account explicitly the free 
propagation of the mode $a$ between DPA and NPA, since this only changes
the phase angle of squeezing. 

The input state $|\psi\rangle_{\rm in}$ satisfies the vacuum condition,
$a |\psi\rangle_{\rm in} = b |\psi\rangle_{\rm in} = 0$, i.e. it is an 
eigenstate of the annihilation operators. Using this property, we find 
that the output state satisfies the equations
\begin{eqnarray}
& & U a U^{\dagger} |\psi\rangle = 
( \mu_1 \mu_2 a + \nu_1 \nu_2^{\ast} b - \nu_1 \mu_2 a^{\dagger} 
- \mu_1 \nu_2 b^{\dagger} ) |\psi\rangle = 0  \label{eeq1a}  \\
& & U b U^{\dagger} |\psi\rangle =
( \mu_2 b - \nu_2 a^{\dagger} ) |\psi\rangle = 0 \label{eeq1b}
\end{eqnarray}
where we have introduced the following notation
\begin{equation}
\xi_j = - {\rm i} g_j t_j = |\xi_j|\, e^{ {\rm i} \theta_j } 
\hspace*{0.8cm}
\mu_j = \cosh |\xi_j| \hspace*{0.8cm} 
\nu_j = \sinh |\xi_j|\, e^{ {\rm i} \theta_j } 
\hspace*{0.8cm} j=1,2. 
\end{equation}

We next assume that a measurement of the number of photons is performed
in one of the output beams by an ideal photodetector with perfect 
quantum efficiency. The outcome of the measurement is denoted by $n$. 
The state in the remaining mode after this measurement is given by
the projection of the number state $|n\rangle_i$ on $|\psi\rangle$,
\begin{equation}
|\psi_n\rangle = {}_i \langle n|\psi \rangle 
\end{equation}
where $i$ is either $a$ or $b$, depending in which mode the measurement
has been performed. 

We first consider the case of the photon-number measurement in the mode
$b$. The reduced eigenvalue equation for the state $|\psi_n\rangle$
can be obtained from equations (\ref{eeq1a}) and (\ref{eeq1b}) by 
rewriting them in the form 
\begin{equation}
b^{\dagger} b |\psi\rangle = ( a^{\dagger} a - \chi a^{\dagger 2} ) 
|\psi\rangle \label{bb}   
\end{equation}
where we have defined 
\begin{equation}
\chi = \frac{\nu_1}{\mu_1 \mu_2^2} = 
\frac{\tanh |\xi_1|}{\cosh^2 |\xi_2|}\, e^{ {\rm i} \theta_1 } \, .
\label{chi-def1}
\end{equation}
Note that $|\chi| < 1$. The eigenvalue equation for $|\psi_n\rangle$ 
is obtained now by projecting (\ref{bb}) over the number state 
$|n\rangle_b$. This gives
\begin{equation}
( a^{\dagger} a - \chi a^{\dagger 2} ) |\psi_n\rangle 
= n |\psi_n\rangle \, . \label{zplus_a}   
\end{equation}

In terms of the SU(1,1) generators, equation (\ref{zplus_a}) can be 
written as
\begin{equation}
( K_{3} - \chi K_{+} ) |\psi_n\rangle = (k+l) |\psi_n\rangle 
\end{equation}
where $l$ is a non-negative integer defined by
\begin{equation}
l = \left[ \frac{n}{2} \right] = \left\{ \begin{array}{l}
n/2 \hspace{0.4cm} \mbox{for $n$ even}\; (k=1/4)  \\
(n-1)/2 \hspace{0.4cm} \mbox{for $n$ odd}\; (k=3/4) \, . \end{array}
\right.   \label{ldef}
\end{equation}
According to this definition, $n=2l+2k-\frac{1}{2}$. We see that the 
state $|\psi_n\rangle$ belongs to the even representation
($k=1/4$) for even $n$, and to the odd representation ($k=3/4$) for odd 
$n$. We also see that in the limit $\chi \rightarrow 0$ (i.e.
$\xi_1 \rightarrow 0$) the state $|\psi_n\rangle$ approaches the number 
state $|n\rangle$ (an eigenstate of $K_{3}$). This behavior is easily 
understood by recalling that for $\xi_1 = 0$ the mode $a$ enters
NPA in the vacuum state. Thus we recover the particular case in which
the number states are generated. 

Let us now consider the effect of performing an additional 
transformation on the mode $a$.
We assume that this mode, after it has been prepared in the state 
$|\psi_n\rangle$, accumulates a phase shift $\varphi$ and is then once 
again squeezed in a lossless degenerate parametric amplifier with a
classical coherent pump. Both these processes can be represented as 
SU(1,1) transformations, and the final state $|\bar{\psi}_n\rangle$ is
given by
\begin{equation}
|\bar{\psi}_n\rangle = \exp( - {\rm i} \omega K_{2} ) 
\exp( - {\rm i} \varphi K_{3} )
|\psi_n\rangle \, .     
\end{equation}
If we choose 
\begin{equation}
\varphi = {\rm arg}\, \chi = \theta_1 \hspace{0.8cm}
\tanh \omega = |\chi|,
\end{equation}
the final state $|\bar{\psi}_n\rangle$ will satisfy the eigenvalue
equation
\begin{equation}
( \eta K_{2} + {\rm i} K_{3} ) |\bar{\psi}_n\rangle = 
{\rm i} (k+l) \sqrt{\eta^2 + 1}\,
|\bar{\psi}_n\rangle \, . \label{ist_yz}
\end{equation}
The real parameter $\eta$ is defined as
\begin{equation}
\eta = \sinh \omega = \frac{|\chi|}{\sqrt{1-|\chi|^2}} \, .
\label{eta_yz}
\end{equation}
The states $|\bar{\psi}_n\rangle$ of equation (\ref{ist_yz}) are 
easily recognized as the $K_{2}$-$K_{3}$ intelligent states. 

We next consider the case of the photon-number measurement in the mode 
$a$. It turns out that this situation contains more possibilities than 
the previous one. The reduced eigenvalue equation for the state 
$|\psi_n\rangle$ is obtained from equations (\ref{eeq1a}) and 
(\ref{eeq1b}) by rewriting them in the form 
\begin{equation}
a^{\dagger} a |\psi\rangle = ( b^{\dagger} b + \chi b^2 ) 
|\psi\rangle  \label{aa}
\end{equation}
where we have defined 
\begin{equation}
\chi = \frac{\nu_1}{\mu_1 \nu_2^2} = \frac{\tanh |\xi_1|
}{\sinh^2 |\xi_2|}\, e^{ {\rm i} (\theta_1 - 2\theta_2 )} \, .
\label{chi-def2}
\end{equation}
Note that here $0 < |\chi| < \infty$. The eigenvalue equation for 
$|\psi_n\rangle$ is obtained now by projecting (\ref{aa}) over the 
number state $|n\rangle_a$. This gives
\begin{equation}
( b^{\dagger} b + \chi b^2 ) |\psi_n\rangle 
= n |\psi_n\rangle \, . \label{zplus_b}   
\end{equation}
In terms of the SU(1,1) generators, this equation takes the form
\begin{equation}
( K_{3} + \chi K_{-} ) |\psi_n\rangle = (k+l) |\psi_n\rangle 
\end{equation}
where $l$ is a non-negative integer defined by equation (\ref{ldef}).
Once again, the state $|\psi_n\rangle$ belongs to the even representation
($k=1/4$) for even $n$, and to the odd representation ($k=3/4$) for odd 
$n$. For $\chi \rightarrow 0$ (i.e. $\xi_1 \rightarrow 0$), we again
go back to the situation where both modes entering NPA are in the vacuum 
state, and the photon-number measurement in one of the output modes leads 
to the number state in the other.

Also, we consider the effect of additional SU(1,1) transformations
performed on the  mode $b$, after it has been prepared in the state 
$|\psi_n\rangle$.
The final state $|\bar{\psi}_n\rangle$ is given by
\begin{equation}
|\bar{\psi}_n\rangle = \exp( {\rm i} \omega K_{2} ) 
\exp( {\rm i} \varphi K_{3} ) |\psi_n\rangle \, .     
\end{equation}
It is convenient to choose 
\begin{equation}
\varphi = {\rm arg}\, \chi = \theta_1 - 2 \theta_2 \, . 
\end{equation}
It is necessary to distinguish here between the two possibilities: 
$|\chi| < 1$ and $|\chi| > 1$. When $|\chi| < 1$, we choose 
\begin{equation}
\tanh \omega = |\chi|,
\end{equation}
which leads to the eigenvalue equation (\ref{ist_yz}) with
the real parameter $\eta$ given by (\ref{eta_yz}) but with $\chi$
of equation (\ref{chi-def2}). In this case we once again obtain the 
$K_{2}$-$K_{3}$ intelligent states.

The case $|\chi| > 1$ is different. Here we choose 
\begin{equation}
\coth \omega = |\chi|,
\end{equation}
which leads to the eigenvalue equation
\begin{equation}
( \eta K_{1} - {\rm i} K_{2} ) |\bar{\psi}_n\rangle = (k+l) 
\sqrt{1 - \eta^2}\, |\bar{\psi}_n\rangle \, . \label{ist_xy}
\end{equation}
The real parameter $\eta$ is defined as
\begin{equation}
\eta = \frac{1}{\cosh \omega} = \frac{\sqrt{|\chi|^2 -1}}{|\chi|} \, .
\end{equation}
Note that here $0< \eta <1$.
The states $|\bar{\psi}_n\rangle$ of equation (\ref{ist_xy}) are  
recognized as the $K_{1}$-$K_{2}$ intelligent states.

\section{Analytic representations}
\label{sec:analyt}
\setcounter{equation}{0}

In this section we present an analytic formalism that provides a 
complete solution to the eigenvalue problem involving any linear
combination of the SU(1,1) generators \cite{Brif:pre}. 
In partcular, we are able to solve the eigenvalue equations that 
define the SU(1,1) intelligent states.
For the two-photon realization of SU(1,1), we use a special treatment 
involving a ``weak'' resolution of the identity in terms of the 
squeezed states \cite{BVM96}.

\subsection{The analytic representation in the unit disk and its 
``weak'' extension}

As was discussed by Perelomov \cite{Per},
each SU(1,1) coherent state corresponds to a point in the coset
space SU(1,1)/U(1) that is the upper sheet of the two-sheet 
hyperboloid (Lobachevski plane). Thus a coherent state is specified 
by a pseudo-Euclidean unit vector of the form 
$(\sinh\tau \cos\varphi , \sinh\tau \sin\varphi , \cosh\tau)$.
The coherent states $| \zeta,k \rangle$ are obtained by applying 
unitary operators 
$\Omega(\xi) \in$ SU(1,1)/U(1) to the lowest state $|n=0,k \rangle$,
        \begin{eqnarray}
| \zeta,k \rangle & = & \exp \left( \xi K_{+} -\, \xi^{\ast} 
K_{-} \right) | 0,k \rangle  = (1-|\zeta|^{2})^{k} \exp 
\left( \zeta K_{+} \right) |0,k \rangle   \nonumber \\
& = & (1-|\zeta|^{2})^{k} \sum_{n = 0}^{\infty} 
\left[\frac{\Gamma(n+2k)}{n!\Gamma(2k)}\right]^{1/2}\!\zeta^{n} 
|n,k \rangle \, .         \label{an.1.1}
        \end{eqnarray}
Here $\xi=-(\tau/2)\, {\rm e}^{- {\rm i} \varphi}$ and 
$\zeta = (\xi/|\xi|) \tanh |\xi| = -\tanh (\tau/2)\, 
{\rm e}^{- {\rm i} \varphi}$, so $|\zeta| < 1$. 
The condition $|\zeta|<1$ shows that the 
SU(1,1) coherent states are defined in the interior 
of the unit disk.
An important property is the resolution of the identity:
for $k>1/2$ one gets
        \begin{equation}
\int {\rm d} \mu (\zeta,k)  | \zeta,k \rangle \langle \zeta,k | = I   
\mbox{\hspace{1.5cm}}
{\rm d} \mu (\zeta,k) = \frac{2k-1}{\pi} \frac{ d^{2} 
\zeta}{(1-|\zeta|^{2})^{2}}        \label{an.1.2}
        \end{equation}
where the integration is over the unit disk $|\zeta|<1$. 
For $k=\frac{1}{2}$ the limit $k\rightarrow \frac{1}{2}$ must be 
taken after the integration is carried out in the general form. 
One can represent the state space ${\cal H}_{k}$ as the Hilbert 
space of analytic functions $G(\zeta;k)$ in the unit disk 
${\cal D}(|\zeta|<1)$. They form the so-called Hardy space 
${\cal H}({\cal D})$. For a normalized state
$| \psi \rangle = \sum_{n = 0}^{\infty} C_{n} |n,k \rangle$, 
one gets 
        \begin{equation}
G(\zeta;k) = (1-|\zeta|^{2})^{-k} \langle \zeta^{\ast},k |\psi\rangle 
= \sum_{n = 0}^{\infty} C_{n} \left[\frac{\Gamma(n+2k)
}{n!\Gamma(2k)}\right]^{1/2}\!\zeta^{n}   
\label{an.1.3}
        \end{equation}
        \begin{equation}
| \psi \rangle = \int {\rm d} \mu (\zeta,k)  (1-|\zeta|^{2})^{k} 
G(\zeta^{\ast};k) | \zeta,k \rangle   \label{an.1.4}
        \end{equation}
and the scalar product is 
        \begin{equation}
\langle \psi_{1} | \psi_{2} \rangle = \int {\rm d} \mu (\zeta,k)  
(1-|\zeta|^{2})^{2k} [G_{1}(\zeta;k)]^{\ast} G_{2}(\zeta;k) \, .
\label{an.1.5}
        \end{equation}
This is the analytic representation in the unit disk.

A serious problem arises for $k < 1/2$, when the resolution of the 
identity (\ref{an.1.2}) does not hold. However, another resolution of 
the identity can be constructed, which is valid for both $k > 1/2$ and 
$k < 1/2$ \cite{BVM96}. We point out that in equation (\ref{an.1.3})
many functions converge in a disk that is larger than the unit disk.
We call ${\cal H}({\cal D}(1+\epsilon))$ the subspace of the Hardy 
space that contains all the functions that converge in the disk
${\cal D}(1+\epsilon) = \{|\zeta|< 1+\epsilon \}$ (where $\epsilon 
> 0$). Clearly if $\epsilon_1 > \epsilon_2$ then
${\cal H}({\cal D}(1+\epsilon_1))$ is a subspace of 
${\cal H}({\cal D}(1+\epsilon_2))$. As $\epsilon$ goes to 0 (from 
above), ${\cal H}({\cal D}(1+\epsilon))$ becomes the Hardy space.
It can be shown \cite{BVM96} that for any positive $k$ apart from
integers and half-integers and for any two states in
$H({\cal D}(1+\epsilon))$ (where $\epsilon$ is any positive number), 
the scalar product can be written in the form 
        \begin{equation}
\langle \psi_1 | \psi_2 \rangle =
\frac{-(2k-1)\exp(2\pi {\rm i} k)}{4\pi {\rm i} \sin(2\pi k)}\,
\oint_{{\cal C}}  \frac{ {\rm d} t}{(1-t)^{2-2k}} \int_{0}^{2\pi} 
d \phi \, [G_{1}(\zeta;k)]^{\ast} G_{2}(\zeta;k)  \label{an.1.6}
        \end{equation}
where $\zeta = \sqrt{t}\exp( {\rm i} \phi)$.
The contour ${\cal C}$ is a single loop that goes from the origin 
up to one below the real axis, turns back around the point $t=1$ in 
the counter-clockwise direction, and goes back above the real axis
up to zero. This contour goes around 1 but is entirely within 
${\cal D}(1+\epsilon)$. 
Equation (\ref{an.1.6}) gives a ``weak'' resolution of the identity 
which we express as 
        \begin{equation}
I = \frac{-(2k-1)\exp(2\pi {\rm i} k)}{4\pi {\rm i} \sin(2\pi k)}
\oint_{{\cal C}} \frac{ {\rm d} t}{(1-t)^{2}} \int_{0}^{2\pi} {\rm d} 
\phi \, |\zeta,k \rangle\langle \zeta,k| \, .   \label{an.1.7}
        \end{equation} 
Although the contour ${\cal C}$ goes outside the unit disk, where the 
SU(1,1) coherent states are not normalisable, this equation has to be 
understood in a weak sense in conjuction with equation (\ref{an.1.6}).
The analytic functions $G(\zeta;k) \in H({\cal D}(1+\epsilon))$ are 
defined according to (\ref{an.1.3}), but now
        \begin{equation}
| \psi \rangle = \frac{-(2k-1)\exp(2\pi {\rm i} k)}{4\pi {\rm i} 
\sin(2\pi k)} \oint_{{\cal C}} \frac{ {\rm d} t}{(1-t)^{2-k}} 
\int_{0}^{2\pi} {\rm d} \phi \, G(\zeta^{\ast};k) |\zeta,k \rangle    
\label{an.1.8}
        \end{equation} 
and the scalar product is given by (\ref{an.1.6}). Therefore,
equations (\ref{an.1.3}), (\ref{an.1.6}) and (\ref{an.1.8}) define
the analytic representation in ${\cal D}(1+\epsilon)$.

We can use these results in the context of the two-photon realization,
where $k=1/4$ and $k=3/4$. The unitary group operator $\Omega(\xi) 
\in$ SU(1,1)/U(1) for the two-photon realization is the well-known 
squeezing operator $S(\xi)$ \cite{Stoler,Yuen}:
        \begin{equation}
S(\xi) = \exp\left( \xi K_{+} - \xi^{\ast} K_{-} \right) =
\exp\left( \frac{\xi}{2} a^{\dagger 2}  - \frac{\xi^{\ast}}{2} a^{2} 
\right) \, .     \label{an.1.9}
        \end{equation}
Therefore, the SU(1,1) coherent states are the squeezed states. 
For $k=1/4$ one gets the squeezed vacuum,
        \begin{equation}
| \zeta,\mbox{$\frac{1}{4}$} \rangle = S(\xi) 
|0\rangle = (1-|\zeta|^{2})^{1/4} \sum_{n=0}^{\infty} 
\frac{ \sqrt{(2n)!} }{ 2^{n} n!} \zeta^{n} |2n\rangle  \label{an.1.10}
        \end{equation}
while for $k=3/4$ one gets the squeezed ``one photon'' state,
        \begin{equation}
| \zeta,\mbox{$\frac{3}{4}$} \rangle = S(\xi) 
|1\rangle = (1-|\zeta|^{2})^{3/4} \sum_{n=0}^{\infty} \frac{ 
\sqrt{(2n+1)!}}{2^{n} n!} \zeta^{n} |2n+1\rangle \, .  \label{an.1.11}
        \end{equation}
As before, $\zeta = (\xi/|\xi|) \tanh |\xi|$. It is also possible to
define the parity-dependent squeezing operator \cite{BMV96} that 
imposes different squeezing transformations on the even and odd 
subspaces of the Fock space.
Using the squeezed states, we obtain the following resolutions of the 
identity (in a weak sense, as explained above):
        \begin{eqnarray}
& & \frac{1}{8\pi} \oint_{{\cal C}} \frac{ {\rm d} t}{(1-t)^{2}} 
\int_{0}^{2\pi} {\rm d} \phi \, | \zeta,\mbox{$\frac{1}{4}$}
\rangle \langle \zeta,\mbox{$\frac{1}{4}$} | = I_{e}
= \sum_{n=0}^{\infty} |2n \rangle\langle 2n| \label{an.1.12} \\
& & -\frac{1}{8\pi} \oint_{{\cal C}} \frac{ {\rm d} t}{(1-t)^{2}} 
\int_{0}^{2\pi} {\rm d} \phi \, | \zeta,\mbox{$\frac{3}{4}$} 
\rangle \langle \zeta,\mbox{$\frac{3}{4}$} | = I_{o}
= \sum_{n=0}^{\infty} |2n+1 \rangle\langle 2n+1| \, . \label{an.1.13}
        \end{eqnarray}
The corresponding analytic representation is given by equations
(\ref{an.1.3}), (\ref{an.1.6}) and (\ref{an.1.8}) with $k=1/4$ and
$k=3/4$ for states in the even and odd subspaces, respectively.
This analytic representation is related to the well-known Bargmann
representation through a Laplace transform (see \cite{BVM96} for more
details).

\subsection{The general eigenvalue problem}

Let us consider a linear combination of the SU(1,1) generators
of the form 
        \begin{equation}
(\vec{\beta} \cdot \vec{K}) = 
\beta_{1} K_{1} +\beta_{2} K_{2} + \beta_{3} K_{3}  \label{an.2.1}
        \end{equation}
where $\beta_{1}$, $\beta_{2}$ and $\beta_{3}$ are complex parameters.
Thus, the operator $(\vec{\beta} \cdot \vec{K})$ belongs to the 
complexified SU(1,1) algebra. The general eigenvalue problem for the 
SU(1,1) group can be expressed as \cite{Brif:pre,Trif:pre1}:
        \begin{equation}
(\vec{\beta}\cdot \vec{K}) |\psi\rangle = 
(\beta_{1} K_{1} +\beta_{2} K_{2} + \beta_{3} K_{3}) 
|\psi\rangle = \lambda |\psi\rangle \hspace{0.8cm} 
|\psi\rangle \in {\cal H}_{k} \, .  
\label{an.2.2}
        \end{equation}
Normalized states $|\psi\rangle$ that satisfy this equation are
called the SU(1,1) algebra eigenstates \cite{Brif:pre} or the SU(1,1)
algebraic coherent states \cite{Trif:pre1}. Many particular cases of 
equation (\ref{an.2.2}) have been considered in literature.
However, the complete solution of the general eigenvalue problem
(\ref{an.2.2}) has been derived only recently \cite{Brif:pre},
using the analytic representation in the unit disk. We recapitulate
here some basic results that are relevant to the single-mode SU(1,1) 
intelligent states, whose generation we have discussed in section
\ref{sec:gener}. 
According to equation (\ref{an.1.3}), the state $|\psi\rangle$ is
represented by the analytic function $G(\zeta;k)$. The SU(1,1) 
generators act in the Hilbert space of analytic functions $G(\zeta;k)$ 
as first-order differential operators: 
        \begin{equation}
K_{+} = \zeta^{2} \frac{d}{d\zeta} + 2k\zeta   
\hspace{0.8cm} K_{-} = \frac{d}{d\zeta} 
\hspace{0.8cm} K_{3} = \zeta \frac{d}{d\zeta} +k \, .    
\label{an.2.3}
        \end{equation}
Therefore, the eigenvalue equation (\ref{an.2.2}) is transformed into 
the first-order linear homogeneous differential equation:
        \begin{equation}
( \beta_{+} + \beta_{3} \zeta + \beta_{-} \zeta^{2} ) 
\frac{d}{d\zeta} G(\zeta;k) + 
( 2k \beta_{-} \zeta + k \beta_{3} -\lambda ) G(\zeta;k) = 0     
\label{an.2.4}
        \end{equation}
where we have defined
$\beta_{\pm} = ( \beta_{1} \pm {\rm i} \beta_{2})/2$. 
Let us also define
        \begin{equation}
B = \sqrt{ \beta_{3}^{2}-\beta_{1}^{2}-\beta_{2}^{2} }  \, .  
\label{an.2.5}
        \end{equation}
Note that $B^{2} = (X,X)$ where $X = (\vec{\beta} \cdot \vec{K})$ 
and $(,)$ is the Killing form \cite{BaRa}. All the operators, whose 
eigenstates can be produced in the scheme of section \ref{sec:gener}, 
are semi-simple elements of the complexified algebra, i.e. their
Killing form is non-zero. 

For $B \neq 0$ and $\beta_{+} \neq 0$, the solution of equation 
(\ref{an.2.4}) is  
        \begin{equation}
G(\zeta;k) = {\cal N}^{-1/2} 
( 1 + \tau_{-} \zeta )^{-k+r} ( 1 + \tau_{+} \zeta)^{-k-r}  
\label{an.2.6}  
        \end{equation}
where ${\cal N}$ is a normalization factor, and we use the
following notation:
        \begin{eqnarray}
\label{an.2.7}
& & \tau_{\pm} = (\beta_{1} - {\rm i} \beta_{2})/(\beta_{3} \pm B)  \\
& & r = \lambda/B  \, .  \label{an.2.8} 
        \end{eqnarray}
Admissible values of $\vec{\beta}$ and $\lambda$ are determined by
the requirement that the function $G(\zeta;k)$ must be analytic in 
the unit disk ${\cal D}$ or, for the ``weak'' case, in the larger 
disk ${\cal D}(1+\epsilon)$. If $|\tau_{+}| < (1+\epsilon)^{-1}$ and 
$|\tau_{-}| < (1+\epsilon)^{-1}$, then there are no restrictions on 
$\lambda$ (i.e. the corresponding elements of the complexified 
algebra have a continuous complex spectrum). If $|\tau_{+}| < 
(1+\epsilon)^{-1}$ and $|\tau_{-}| \geq (1+\epsilon)^{-1}$, then the 
analyticity condition requires $r = k+l$ (where $l=0,1,2,\ldots$), 
i.e. the spectrum is discrete and equidistant:
        \begin{equation}
\lambda = (k+l)B \, .   \label{an.2.9}
        \end{equation}
If $|\tau_{+}| \geq (1+\epsilon)^{-1}$ and $|\tau_{-}| < 
(1+\epsilon)^{-1}$, then the analyticity condition requires 
$r = -(k+l)$, and once again the spectrum is discrete and equidistant:
        \begin{equation}
\lambda = -(k+l)B \, .   \label{an.2.10}
        \end{equation}
If $|\tau_{+}| \geq (1+\epsilon)^{-1}$ and $|\tau_{-}| \geq 
(1+\epsilon)^{-1}$, then the function $G(\zeta;k)$ of equation 
(\ref{an.2.6}) cannot be analytic in the disk ${\cal D}(1+\epsilon)$ 
for any value of $\lambda$. This region in the parameter space is 
forbidden, i.e. the corresponding elements of the complexified 
algebra have no normalizable eigenstates. According to equations
(\ref{ist_yz}) and (\ref{ist_xy}), the SU(1,1) intelligent states 
generated in our scheme belong to a class of the algebra eigenstates
with the discrete spectrum. 
For $B \neq 0$ and $\beta_{+} = 0$, the solution is
        \begin{equation}
G(\zeta;k) = {\cal N}^{-1/2} 
\zeta^{l} (1 + \tau_{+} \zeta )^{-2k-l} \label{an.2.11}
        \end{equation}
where $\tau_{+} = \beta_{1}/\beta_{3}$ and $l = -k+\lambda/\beta_{3}$.
The condition of the analyticity requires $l = 0,1,2,\ldots$
(i.e. the spectrum $\lambda = (k+l)\beta_{3}$ is discrete) and 
$|\tau_{+}| < (1+\epsilon)^{-1}$. The case of the vanishing Killing 
form $B = 0$ (the so-called degenerate case) has been also discussed
in \cite{Brif:pre}.

\section{Quantum statistical properties}
\label{sec:prop}
\setcounter{equation}{0}

\subsection{Photon statistics and squeezing}

Let us consider a normalized state $|\psi\rangle = 
\sum_{n=0}^{\infty} C_{n} |n,k\rangle$ that belongs to the Hilbert 
space ${\cal H}_{k}$ of a unitary irreducible representation of 
SU(1,1). For the two-photon realization of SU(1,1), $|\psi\rangle$ 
belongs to the even Fock subspace ${\cal H}_{e}$ when $k=1/4$, and
to the odd Fock subspace ${\cal H}_{o}$ when $k=3/4$. 
The photon-number distribution is $P(m) = |\langle m | \psi 
\rangle|^{2}$, where $|m\rangle$ is a Fock state (photon-number
eigenstate). For $k=1/4$, $P(m)$ does not vanish for even $m$ only: 
$P(2n) = |C_{n}|^{2}$; for $k=3/4$, $P(m)$ does not vanish 
for odd $m$ only: $P(2n+1) = |C_{n}|^{2}$.
The photon-number operator $N = a^{\dagger} a$ can be written as
$N = 2 K_{3} - \frac{1}{2}$. Therefore, the mean photon number and
the variance are given by
        \begin{equation}
\langle N \rangle = 2 \langle K_{3} \rangle - \mbox{$\frac{1}{2}$}
\hspace{0.8cm} (\Delta N)^{2} = 4 (\Delta K_{3})^{2} \, .
\label{p.1.1}
        \end{equation}
Photon statistics can be conveniently characterized by the 
intensity correlation function:
        \begin{equation}
g^{(2)} = 1 + \frac{ (\Delta N)^{2} - \langle N \rangle }{
\langle N \rangle^{2} } \, .
\label{p.1.2}
        \end{equation}
Photon statistics is sub- or super-Poissonian for $g^{(2)} <1$ or
$g^{(2)} >1$, respectively. The minimal available value of $g^{(2)}$
is zero, corresponding to the maximal possible photon antibunching.

If $A$ and $B$ are two non-commuting observables, $[A,B] = {\rm i} C$,
the product $(\Delta A)^{2} (\Delta B)^{2}$ taken over a quantum
state $|\psi\rangle$ must satisfy the uncertainty relation 
(\ref{SR-unrel}). One of the most intriguing phenomena in quantum 
optics is squeezing, when the quantum noise in one observable is 
reduced on account of its counter-partner. The state $|\psi\rangle$
is called squeezed in $A$ or $B$, if 
        \begin{equation}
(\Delta A)^{2} < \mbox{$\frac{1}{2}$} |\langle C \rangle| 
\;\;\; {\rm or} \;\;\; 
(\Delta B)^{2} < \mbox{$\frac{1}{2}$} |\langle C \rangle| \, .
\label{p.1.3}
        \end{equation}
It is clear that the ordinary intelligent state, for which
$(\Delta A)^{2} (\Delta B)^{2} = \frac{1}{4} \langle C \rangle^{2}$,
is squeezed in either $A$ or $B$ whenever the two uncertainties
are unequal. Note that condition
(\ref{p.1.3}) determines a relation between uncertainties of $A$ and 
$B$. However, the uncertainty of the squeezed observable can be quite
large if $\frac{1}{2} |\langle C \rangle|$ is large. Therefore, one
should define a more restrictive condition of squeezing 
\cite{Trif,Trif:pre2}:
        \begin{equation}
(\Delta A)^{2} < \Delta_{0}^{2} \;\;\; {\rm or} \;\;\; 
(\Delta B)^{2} < \Delta_{0}^{2} \hspace{0.8cm} 
\Delta_{0}^{2} = \min \left( \mbox{$\frac{1}{2}$} 
|\langle C \rangle| \right) \, .       \label{p.1.4}
        \end{equation}
If condition (\ref{p.1.3}) or (\ref{p.1.4}) is satisfied, we refer to 
this phenomenon as {\em relative\/} or {\em absolute\/} squeezing,
respectively. Obviously, if a state is absolutely squeezed, it is 
relatively squeezed too. However, this relation is not valid in the 
opposite direction. 

Usual (linear) squeezing is defined for the field quadratures
$q$ and $p$, which are given by
        \begin{equation}
q = \frac{ a + a^{\dagger} }{ \sqrt{2} } \hspace{0.8cm} 
p = \frac{ a - a^{\dagger} }{ {\rm i} \sqrt{2} } \hspace{0.8cm} 
[q,p] = {\rm i} \, .  \label{p.1.5}
        \end{equation}
Since $q$ and $p$ are canonically conjugate (i.e. their commutator
is a $c$-number), there is no difference between relative and 
absolute squeezing. The state $|\psi\rangle$ is called linearly 
squeezed if
        \begin{equation}
(\Delta q)^{2} < \mbox{$\frac{1}{2}$} \;\;\; {\rm or} \;\;\; 
(\Delta p)^{2} < \mbox{$\frac{1}{2}$} \, .  \label{p.1.6}
        \end{equation}
For photon states that belong to either the even or the odd Fock 
subspace (i.e. to an irreducible representation of SU(1,1) with 
$k=1/4$ or $k=3/4$), the mean values of the field quadratures vanish:
        \begin{equation}
\langle q \rangle = \langle p \rangle = 0 \, .   \label{p.1.7}
        \end{equation}
Let us define $Q_{+} = q$ and $Q_{-} = p$. Then one obtains
        \begin{equation}
(\Delta Q_{\pm})^{2} = \langle Q_{\pm}^{2} \rangle = 
\mbox{$\frac{1}{2}$} \langle a^{\dagger} a + a a^{\dagger} 
\pm a^{2} \pm a^{\dagger 2} \rangle = 
2 ( \langle K_{3} \rangle \pm \langle K_{1} \rangle ) .  
\label{p.1.8} 
        \end{equation}

Quadratic squeezing is defined (see e.g. \cite{BHY_YH}) for the 
quadrature components of the squared annihilation operator $a^{2}$. 
In the context of the two-photon realization, quadratic squeezing is 
equivalent to squeezing in the generators $K_{1}$ or $K_{2}$ of 
SU(1,1). Both linear and squared quadratures $q$, $p$ and $K_{1}$, 
$K_{2}$ have no normalizable eigenstates. Squeezing of these 
observables is interesting, because their variances never vanish 
exactly. The condition for relative quadratic squeezing can be 
expressed in the form
        \begin{equation}
(\Delta K_{1})^{2} < \mbox{$\frac{1}{2}$} \langle K_{3} \rangle 
\;\;\; {\rm or} \;\;\; 
(\Delta K_{2})^{2} < \mbox{$\frac{1}{2}$} \langle K_{3} \rangle \, .
\label{p.1.9}
        \end{equation}
Since $\Delta_{0}^{2} = \min (\frac{1}{2} \langle K_{3} \rangle)
= k/2$, the condition for absolute quadratic squeezing is
        \begin{equation}
(\Delta K_{1})^{2} < k/2 \;\;\; {\rm or} \;\;\; 
(\Delta K_{2})^{2} < k/2 \, .     \label{p.1.10}
        \end{equation}
Note that $\Delta_{0}^{2}$ is equal to $1/8$ and $3/8$ for the even
and the odd Fock subspaces, respectively. The SU(1,1) coherent states 
for the two-photon realization are the usual (linear) squeezed states.
These states can also exhibit relative quadratic squeezing (e.g.
states which are simultaneously coherent and intelligent).
However, absolute quadratic squeezing cannot be achieved by any
SU(1,1) coherent state. Nevertheless, there exist states which 
exhibit simultaneously both linear and absolute quadratic squeezing 
\cite{Trif:pre2}.

\subsection{General results}

The decomposition of a normalized state $|\psi\rangle$ over the
orthonormal basis $|n,k\rangle$ can be obtained by expanding its
analytic function $G(\zeta;k)$ into a power series in $\zeta$. The 
function $G(\zeta;k)$ of equation (\ref{an.2.6}) is a generating 
function for the Jacobi polynomials, and its power-series expansion 
is \cite{Brif:pre}
        \begin{equation}
G(\zeta;k) = {\cal N}^{-1/2} \sum_{n=0}^{\infty} 
P_{n}^{(-k+r-n,-k-r-n)}(x)\, (\kappa \zeta)^{n}  \label{p.2.1}
        \end{equation}
where
        \begin{eqnarray}
& & \kappa = \tau_{+} - \tau_{-} = -2B/(\beta_{1}+i\beta_{2}) 
\label{p.2.2} \\
& & x = (\tau_{-} + \tau_{+})/(\tau_{-} - \tau_{+}) = \beta_{3}/B \, .
\label{p.2.3}
        \end{eqnarray}
Comparing the expansion (\ref{p.2.1}) with the general formula 
(\ref{an.1.3}), we find the decomposition of the corresponding 
state $|\psi\rangle$ over the orthonormal basis:
        \begin{equation}
|\psi\rangle = {\cal N}^{-1/2} \sum_{n=0}^{\infty}
\left[ \frac{ n! \Gamma(2k) }{ \Gamma(2k+n) } \right]^{1/2}
P_{n}^{(-k+r-n,-k-r-n)}(x)\, \kappa^{n} |n,k\rangle \, .
\label{p.2.4}
        \end{equation}
The normalization factor is given by \cite{Brif:pre}
        \begin{eqnarray}
{\cal N} & = & \sum_{n=0}^{\infty} \frac{ n! \Gamma(2k) }{ 
\Gamma(2k+n) } \left| P_{n}^{(-k+r-n,-k-r-n)}(x) \right|^2 t^{n} 
\nonumber \\
& = & S_{+}^{-k+r} S_{-}^{-k-r}
F\left(k+r,k-r;2k;-\frac{t}{S_{+}S_{-}} \right) \label{p.2.5}
        \end{eqnarray}
where $t = |\kappa|^2$, $r$ is assumed to be real, $F(a,b;c;z)$ is 
the hypergeometric function and
        \begin{equation}
S_{\pm} = 1 - |x \pm 1|^{2} t/4 = 1 - |\tau_{\mp}|^{2} \, .
\label{p.2.6}
\label{Sk-def}
        \end{equation}
In the case of the discrete spectrum (i.e. $r=k+l$ or $r=-(k+l)$),
one can use the relation between the hypergeometric function and the 
Jacobi polynomials \cite{Erd}, in order to obtain
        \begin{equation}
{\cal N} = \frac{ l! \Gamma(2k) }{ \Gamma(2k+l) } 
S_{i}^{l} S_{i'}^{-2k-l} P_{l}^{(2k-1,0)}
\left( 1+\frac{2t}{S_{+}S_{-}} \right) \, .
\label{p.2.7}
        \end{equation}
Here $(i,i') = (+,-)$ for $r=k+l$ and $(i,i') = (-,+)$ for 
$r=-(k+l)$. 

The above analytic expressions can be used for calculations of 
quantum statistical properties of the corresponding eigenstates.
By using the property $K_{3} |k,n\rangle = (k+n) |k,n\rangle$ 
and formula (\ref{p.2.5}), moments of $K_{3}$ can be
expressed as derivatives of ${\cal N}$ with respect to $t$:
        \begin{equation}
\langle K_{3} \rangle = \frac{t}{{\cal N}} 
\frac{\partial {\cal N}}{\partial t} + k   \hspace{0.8cm}
(\Delta K_{3})^{2} = \frac{t^{2}}{{\cal N}} 
\frac{\partial^{2} {\cal N}}{\partial t^{2}} +
\frac{t}{{\cal N}} \frac{\partial {\cal N}}{\partial t}
- \left( \frac{t}{{\cal N}} \frac{\partial {\cal N}}{\partial t}
\right)^{2} \, . 
\label{p.2.8}
        \end{equation}
By using the formula \cite{Erd}
        \begin{equation}
\frac{ {\rm d} F(a,b;c;z) }{ {\rm d} z } = \frac{ab}{c} F(a+1,b+1;c+1;z)
\label{p.2.9}
        \end{equation}
and the hypergeometric equation, we obtain exact analytic expressions 
for the moments of $K_{3}$ over the state $|\psi\rangle$ of 
equation (\ref{p.2.4}):
        \begin{equation}
\langle K_{3} \rangle = \frac{-k Y + r(S_{+}-S_{-})}{ S_{+}S_{-} }
+ \frac{ (k^{2}-r^{2}) Y t }{ 2 k S_{+}^{2} S_{-}^{2} }\, \Theta 
\label{p.2.10}
        \end{equation}
        \begin{eqnarray}
(\Delta K_{3})^{2} & = & (k+r) \frac{ 1-S_{-} }{ S_{-}^{2} }
+ (k-r) \frac{ 1-S_{+} }{ S_{+}^{2} } - \frac{ (k^{2}-r^{2})
Y^{2} t }{ ( S_{+}S_{-} + t) S_{+}^{2} S_{-}^{2} } \nonumber \\
& & - \frac{ (k^{2}-r^{2})\, t }{ 2 k S_{+}^{3} S_{-}^{3} }
\left( \frac{ S_{+} S_{-} Y^{2} }{ S_{+}S_{-} + t } - 2 k Y^2 
+ Z \right) \Theta  - \frac{ (k^{2}-r^{2})^{2} Y^{2} t^{2} }{ 
4 k^{2} S_{+}^{4} S_{-}^{4} }\, \Theta^{2} \, .
\label{p.2.11}
        \end{eqnarray}
Here, we use the following notation:
        \begin{equation}
Y = S_{+} S_{-} - S_{+} - S_{-} 
\label{p.2.12}
        \end{equation}
        \begin{equation}
Z = S_{+}^{2}(1- S_{-}) + S_{-}^{2}(1 - S_{+})  
\label{p.2.13}
        \end{equation}
        \begin{equation}
\Theta = \left[ F\left(k+r,k-r;2k;-\frac{t}{S_{+}S_{-}} \right) 
\right]^{-1} F\left(k+r+1,k-r+1;2k+1;-\frac{t}{S_{+}S_{-}} 
\right) \, .
\label{p.2.14}
        \end{equation}
For $r = \pm (k+l)$ with $l > 0$, the relation between the 
hypergeometric function and the Jacobi polynomials \cite{Erd} gives
        \begin{equation}
\Theta = \frac{2k}{l} \left[ P_{l}^{(2k-1,0)} 
\left(1+\frac{2t}{S_{+}S_{-}} \right) \right]^{-1} 
P_{l-1}^{(2k,1)} \left(1+\frac{2t}{S_{+}S_{-}} \right) \, .
\label{p.2.15}
        \end{equation}
For $r = \pm k$ (i.e. $l = 0$), we obtain $\Theta = 0$, and then 
we recover the known results for the SU(1,1) coherent states 
\cite{WodEb}. The expressions for $\langle K_{3} \rangle$ and 
$(\Delta K_{3})^{2}$ are significantly simplified in the case 
$Y = 0$, which means 
        \begin{equation}
|\tau_{+}\tau_{-}| = 1 \; \Leftrightarrow \; \left| \frac{
(\beta_{1} - {\rm i} \beta_{2})^2 }{ \beta_{1}^{2} + \beta_{2}^{2} }
\right| = 1 \, .    
\label{p.2.16}
        \end{equation}
This condition is satisfied in the important case 
$\beta_{1} = a \beta_{2}$, where $a$ is any real number
(this includes the case when $\beta_{1}$ or $\beta_{2}$
vanishes). Then we obtain
        \begin{equation}
\langle K_{3} \rangle = \frac{h+1}{h-1}\, r 
\label{p.2.17}
        \end{equation}
        \begin{equation}
(\Delta K_{3})^{2} = \frac{ 2 k h }{ (h-1)^{2} } + 
\frac{ (k^{2}-r^{2}) h^2 t }{ k (h-1)^{4} }\, \Theta 
\label{p.2.18}
        \end{equation}
where $h$ is defined by 
        \begin{equation}
h = |\tau_{-}|^2 = 1/|\tau_{+}|^2 \, .  
\label{p.2.19}
        \end{equation}

For the normalized state $|\psi\rangle$, whose analytic function is 
given by equation (\ref{an.2.11}), the procedure is analogous.
One obtains \cite{Brif:pre}
        \begin{equation}
|\psi\rangle = {\cal N}^{-1/2} \sum_{n=l}^{\infty} 
\left[ \frac{ n! \Gamma(2k+n) }{ l! \Gamma(2k+l) } \right]^{1/2} 
\frac{ (-\tau_{+})^{n-l} }{ (n-l)! } |n,k\rangle 
\label{p.2.20}
        \end{equation}
        \begin{equation}
{\cal N} = F\left( l+1,l+2k;1;|\tau_{+}|^{2} \right)
= (1-|\tau_{+}|^{2})^{-2k-l} \, 
P_{l}^{(0,2k-1)} \left( \frac{ 1 + |\tau_{+}|^{2} }{
1-|\tau_{+}|^{2} } \right) \, .
\label{p.2.21}
        \end{equation}
Then we find the following expressions for the moments of $K_{3}$:
        \begin{equation}
\langle K_{3} \rangle = k + l + (l+1)(l+2k) |\tau_{+}|^{2} \Upsilon
 \label{p.2.22}
        \end{equation}
        \begin{eqnarray}
(\Delta K_{3})^{2} & = & \frac{ (l+1)(l+2k) |\tau_{+}|^{2} }{
1 - |\tau_{+}|^{2} } + \frac{ (l+1)(l+2k)(2l+2k+1) |\tau_{+}|^{4} }{
1 - |\tau_{+}|^{2} } \Upsilon \nonumber \\
& & - (l+1)^{2}(l+2k)^{2} |\tau_{+}|^{4} \Upsilon^{2} 
\label{p.2.23}
        \end{eqnarray}
where we have defined
        \begin{equation}
\Upsilon = \frac{ F\left( l+2,l+2k+1;2;|\tau_{+}|^{2} \right) }{
F\left( l+1,l+2k;1;|\tau_{+}|^{2} \right) } \, .
\label{p.2.24}
        \end{equation}

\subsection{Properties of the intelligent states}

The above general results can be applied to the states, whose 
generation has been discussed in section \ref{sec:gener}. 
We will study photon statistics and squeezing properties of these
states.

\subsubsection{Eigenstates of $K_{3} - \chi K_{+}$} 

Here, $\vec{\beta} = (-\chi,- {\rm i} \chi,1)$, $B = 1$ and 
$\beta_{+} = 0$. The corresponding analytic function is given by 
equation (\ref{an.2.11}) with $\tau_{+} = -\chi$ and the eigenvalues are
discrete: $\lambda = k+l$. According to the definition (\ref{chi-def1}) 
of $\chi$, for any $|\xi_{1}| < \infty$ and any $|\xi_{2}| > 0$ there
exists $\epsilon > 0$ such that $|\tau_{+}| < (1+\epsilon)^{-1}$,
so this function belongs to ${\cal H}({\cal D}(1+\epsilon))$. 
For $l=0$, we obtain the SU(1,1) generalized coherent states, i.e.
the squeezed vacuum $|\zeta_{0},\frac{1}{4}\rangle$ for $k=1/4$ and 
the squeezed ``one photon'' state $|\zeta_{0},\frac{3}{4}\rangle$ for 
$k=3/4$. The corresponding SU(1,1) coherent-state amplitude (i.e. the 
squeezing parameter) is $\zeta_{0} = \chi$.

The moments of the generator $K_{3}$ are given by equations 
(\ref{p.2.22}) and (\ref{p.2.23}) with $\tau_{+} = -\chi$. 
The mean and variance of the photon-number operator are then obtained 
from relation (\ref{p.1.1}). For $l=0$, we have $\Upsilon = 
1/(1-|\chi|^2)$ and we recover the results for the SU(1,1) generalized 
coherent states:
        \begin{equation}
\langle K_{3} \rangle = k \frac{ 1+|\chi|^2 }{ 1-|\chi|^2 } 
\hspace{0.8cm}
(\Delta K_{3})^{2} = \frac{ 2 k |\chi|^2 }{ (1-|\chi|^2)^2 } 
\label{p.3.1.1}
        \end{equation}
        \begin{equation}
g^{(2)} = 1 + 2 \frac{ 1-4k + (16k-2)|\chi|^2 + (4k+1)|\chi|^4 }{
[ 4k-1 + (4k+1)|\chi|^2 ]^2 } \, . \label{p.3.1.2}
        \end{equation}
For $n \geq 1$ and sufficiently small $|\chi|$, we find $g^{(2)} < 1$,
i.e. photon statistics is sub-Poissonian. We would like to study the 
limiting behaviour of photon statistics. For $|\chi| \ll 1$, we obtain
        \begin{eqnarray}
& & \langle K_{3} \rangle \approx k+l + (l+1)(l+2k) |\chi|^{2} 
\label{p.3.1.3} \\
& & (\Delta K_{3})^{2} \approx (l+1)(l+2k) |\chi|^{2} \label{p.3.1.4} \\
& & g^{(2)} \approx 1 - \frac{1}{n} \, . \label{p.3.1.5}
        \end{eqnarray}
The expression for $g^{(2)}$ is valid for $n \geq 1$ and for 
$n = 0$ equation (\ref{p.3.1.2}) gives that $g^{(2)} \rightarrow \infty$ 
as $|\chi| \rightarrow 0$. Note that for $\chi \rightarrow 0$ the 
eigenstates of $K_{3} - \chi K_{+}$ reduce to the Fock states 
$|n\rangle$. The maximal available antibunching $g^{(2)} \rightarrow 0$ 
is obtained for $n=1$ as $|\chi| \rightarrow 0$. 
For $|\chi| \rightarrow 1$, we get
        \begin{eqnarray}
& & \langle K_{3} \rangle \approx k+l + \frac{ 2(l+k) |\chi|^{2} }{ 
1 - |\chi|^{2} }  \label{p.3.1.6} \\
& & (\Delta K_{3})^{2} \approx \frac{ (l+1)(l+2k) |\chi|^{2} }{
1 - |\chi|^{2} }  \label{p.3.1.7} \\
& & g^{(2)} \approx \frac{2n+3}{2n+1} \, . \label{p.3.1.8}
        \end{eqnarray}

By the definition of the eigenstates,
$\langle K_{3} - \chi K_{+} \rangle = k+l$.
Then, using expression (\ref{p.2.22}) for $\langle K_{3} \rangle$, 
we obtain
        \begin{eqnarray}
& & \langle K_{1} \rangle = (l+1)(l+2k) \Upsilon |\chi| \cos \theta_1
\label{p.3.1.10} \\
& & \langle K_{2} \rangle = -(l+1)(l+2k) \Upsilon |\chi| \sin \theta_1
\label{p.3.1.11}
        \end{eqnarray}
where $\theta_1 = \arg \chi$. According to equation (\ref{p.1.8}), we 
find the expressions for the uncertainties of the field quadratures:
        \begin{equation}
(\Delta Q_{\pm})^{2} = 2 (k+l) + 2 (l+1)(l+2k) \Upsilon 
(|\chi|^2 \pm |\chi| \cos \theta_1 )  \label{p.3.1.12} 
        \end{equation}
(recall that $Q_{+} = q$ and $Q_{-} = p$). We see that 
$(\Delta q)^{2}(\theta_{1}) = (\Delta p)^{2}(\theta_{1}+\pi)$.
For $\theta_1 = \pi/2$, both uncertainties are equal. For $\theta_1$
equal to zero or $\pi$, one of the uncertainties is maximized and
the other is minimized as functions of $\theta_1$. If we search for
squeezing in the $q$ quadrature, we should put $\theta_1 = \pi$.
For $l=0$, we recover the known results for the ordinary squeezed
states ($\theta_1 = \pi$):
        \begin{equation}
(\Delta Q_{\pm})^{2} = 2 k \frac{ 1 \mp |\chi| }{ 1 \pm |\chi| } .
\label{p.3.1.13} 
        \end{equation}
Then the uncertainty product is $(\Delta q)^{2}(\Delta p)^{2} = (2k)^2$;
for the squeezed vacuum ($k=1/4$) it is minimized: 
$(\Delta q)^{2}(\Delta p)^{2} = 1/4$. 
In the limit $|\chi| \ll 1$, we have ($\theta_1 = \pi$):
        \begin{equation}
(\Delta Q_{\pm})^{2} \approx 2 (k+l) \mp 2 (l+1)(l+2k) |\chi|  .
\label{p.3.1.14} 
        \end{equation}
In this limit very weak squeezing (disappearing as 
$|\chi| \rightarrow 0$) is possible only for $n=0$
(the squeezed vacuum state). Squeezing appears as $|\chi|$ increases.
In the limit $|\chi| \rightarrow 1$ we obtain ($\theta_1 = \pi$):
        \begin{equation}
(\Delta Q_{\pm})^{2} \approx (n + 1/2) 
\frac{ 1 \mp |\chi| }{ 1 \pm |\chi| } .   \label{p.3.1.15} 
        \end{equation}
We see that squeezing becomes weaker as $n$ increases.

\subsubsection{Eigenstates of $K_{3} + \chi K_{-}$} 

Here, $\vec{\beta} = (\chi,- {\rm i} \chi,1)$, $B = 1$ and 
$\beta_{-} = 0$. The corresponding analytic function is given by 
equation (\ref{an.2.6}) with $\tau_{+} = 0$ and $\tau_{-} = 1/\chi$.
Since here $0 < |\chi| < \infty$, the analyticity is guaranteed by
the discrete spectrum, $\lambda = k+l$. For $l=0$, we obtain the
vacuum state $|0\rangle$ when $k=1/4$, or the ``one photon'' state
$|1\rangle$ when $k=3/4$.

The parameters used in the general expressions are $\kappa = -1/\chi$,
$t = 1/|\chi|^2$, $x = 1$, $S_{+} = 1-t$, $S_{-}=1$, $Y = -1$, and
$Z=t$. Then we obtain
        \begin{equation}
\langle K_{3} \rangle = k - \frac{ lt }{ 1-t } + 
\frac{ l (l+2k) t }{ 2k (1-t)^2 }\, \Theta    \label{p.3.2.1}
        \end{equation}
        \begin{equation}
(\Delta K_{3})^{2} = \frac{ l(l+2k-1) t }{ (1-t)^2 }
+ \frac{ l (l+2k) (1-2k) t }{ 2k (1-t)^3 }\, \Theta
- \frac{ l^2 (l+2k)^2 t^2 }{ (2k)^2 (1-t)^4 }\, \Theta^2 
\label{p.3.2.2}
        \end{equation}
        \begin{equation}
\Theta = \frac{2k}{l} \left[ P_{l}^{(2k-1,0)}\left( \frac{1+t}{1-t}
\right) \right]^{-1} P_{l-1}^{(2k,1)}\left( \frac{1+t}{1-t} \right) 
\;\;\;\;\;\;\; l \neq 0     \label{p.3.2.3} 
        \end{equation}
and $\Theta = 0$ for $l=0$. In what follows we will not consider the
trivial case $l=0$. 
The intensity correlation function $g^{(2)}$ is shown in figure~2 
versus $t$, for different values of $n=2l+2k-\frac{1}{2}$.
In the limit $t \ll 1$ ($|\chi| \gg 1$), we obtain
        \begin{equation}
\langle K_{3} \rangle \approx k + \frac{l^2 t}{2k} \hspace{0.8cm} 
(\Delta K_{3})^{2} \approx \frac{l^2 t}{2k}   \label{p.3.2.4}
        \end{equation}
        \begin{equation}
g^{(2)} \approx \frac{1}{4 l^2 t} \;\; {\rm for}\; k=1/4 \;\;\;\;\;
{\rm and}\;\;\;\;\; g^{(2)} \approx 4 l^2 t  \;\; {\rm for}\; k=3/4 .
\label{p.3.2.5}
        \end{equation}
\begin{figure}[htbp]\vspace*{-3.0cm}
\begin{minipage}[c]{.58\linewidth}
\epsfxsize=0.9\textwidth
\centerline{\epsffile{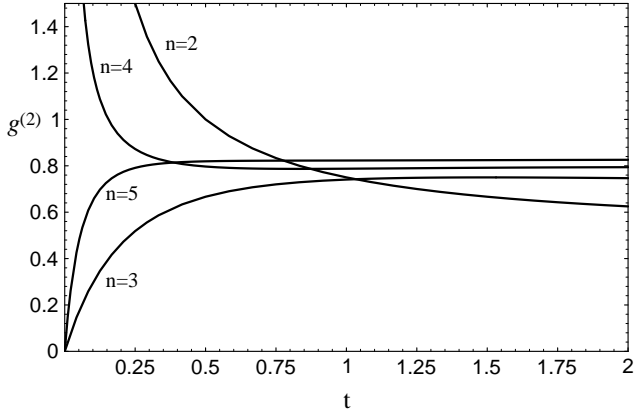}}\vspace*{-2.5cm}
\end{minipage}
\begin{minipage}[c]{.36\linewidth}
\caption{The intensity correlation function $g^{(2)}$ for 
the eigenstates of $K_{3} + \chi K_{-}$ versus $t=1/|\chi|^2$, 
for different values of $n$.}
\label{fig:2}
\end{minipage}
\end{figure}
In this limit photon statistics behaves quite differently for states 
in even and odd subspaces. The smaller $t$ is, the stronger is 
antibunching for odd states and the stronger is bunching for even 
states. Note that in the limit $\chi \rightarrow \infty$ the 
eigenstates of $K_{3} + \chi K_{-}$ become very close to the 
eigenstates of $K_{-}$ (the so-called even and odd coherent states).
In the limit $t \gg 1$ ($|\chi| \ll 1$), we get
        \begin{equation}
\langle K_{3} \rangle \approx k+l - \frac{l(l+2k-1)}{t} \hspace{0.8cm} 
(\Delta K_{3})^{2} \approx \frac{l(l+2k-1)}{t}   \label{p.3.2.6}
        \end{equation}
        \begin{equation}
g^{(2)} \approx \left( 1 - \frac{1}{n} \right) 
\left( 1 + \frac{1}{2t} \right) .  \label{p.3.2.7}
        \end{equation}
As $t$ grows up, $g^{(2)}$ approaches a constant value $1-n^{-1}$.
Note that in the limit $\chi \rightarrow 0$ the eigenstates of 
$K_{3} + \chi K_{-}$ reduce to the Fock states $|n\rangle$. 

By the definition of the eigenstates,
$\langle K_{3} + \chi K_{-} \rangle = k+l$.
Then, using expression (\ref{p.3.2.1}) for $\langle K_{3} \rangle$, 
we obtain
        \begin{equation}
\langle K_{1,2} \rangle = \left[ \frac{l}{1-t} - 
\frac{l(l+2k)t}{2k(1-t)^2} \Theta \right] \sqrt{t} \left\{
\begin{array}{l} \cos\theta \\ \sin\theta \end{array} \right\}
\label{p.3.2.9} 
        \end{equation}
where $\cos\theta$ stands for $\langle K_{1} \rangle$,
$\sin\theta$ for $\langle K_{2} \rangle$, and
$\theta = \arg \chi = \theta_1 - 2\theta_2$. 
According to equation (\ref{p.1.8}), we find the expressions for the 
uncertainties of the field quadratures:
        \begin{equation}
(\Delta Q_{\pm})^{2} = 2 (k+l) - 2 \left[ \frac{l}{1-t} -
\frac{l(l+2k)t}{2k(1-t)^2} \Theta \right] \left( 1 \mp \sqrt{t} 
\cos\theta \right) \, .    \label{p.3.2.10} 
        \end{equation}
We see that $(\Delta q)^{2}(\theta) = 
(\Delta p)^{2}(\theta +\pi)$. For $\theta = \pi/2$, both 
uncertainties are equal. For $\theta$ equal to zero or $\pi$, 
one of the uncertainties is maximized and the other is minimized 
as functions of $\theta$. It can be shown that the expression in the 
square brackets (which is equal to $k+l-\langle K_{3} \rangle$) is
always nonnegative. Therefore, if we search for squeezing in the $q$ 
quadrature, we should put $\theta = \pi$.
The quadrature uncertainty $(\Delta q)^{2}$ is shown in figure~3 
versus $u = 1/|\chi|$, for $\theta = \pi$ and different values of $n$. 
We find that squeezing is possible only for even states ($k=1/4$).  
In the limit $t \ll 1$ ($|\chi| \gg 1$), we have ($\theta = \pi$):
        \begin{equation}
(\Delta Q_{\pm})^{2} \approx 2 k \mp 2 l u + l^2 u^2 / k  
\label{p.3.2.11} 
        \end{equation}
\begin{figure}[htbp]\vspace*{-3.0cm}
\begin{minipage}[c]{.58\linewidth}
\epsfxsize=0.9\textwidth
\centerline{\epsffile{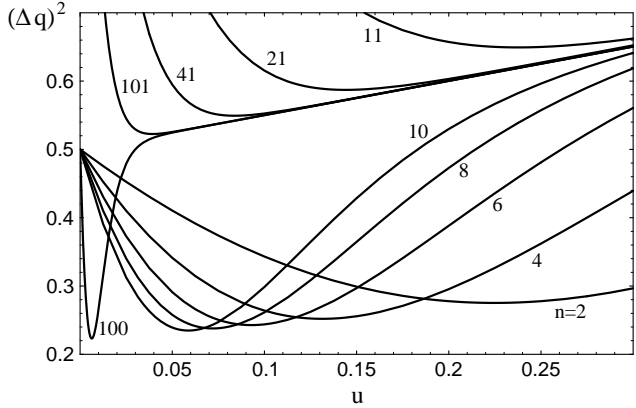}}\vspace*{-2.5cm}
\end{minipage}
\begin{minipage}[c]{.36\linewidth}
\caption{The quadrature uncertainty $(\Delta q)^{2}$ for 
the eigenstates of $K_{3} + \chi K_{-}$ versus $u=1/|\chi|$, 
for $\theta = \pi$ and different values of $n$.}
\label{fig:3}
\end{minipage}
\end{figure}
where $u = \sqrt{t} = 1/|\chi|$. For $\theta = \pi$ and $k=1/4$, the 
uncertainty $(\Delta q)^{2}$ has a minimum ($\approx 0.25$) for $u 
\approx 1/(2n)$. As $n$ increases, this minimum becomes sharper. 
In the limit $t \gg 1$ ($|\chi| \ll 1$) squeezing is impossible:
        \begin{equation}
(\Delta Q_{\pm})^{2} \approx 2(k+l) \mp 2l(l+2k-1) |\chi| .   
\label{p.3.2.12}
        \end{equation}

\subsubsection{Eigenstates of $\eta K_{1} - {\rm i} K_{2}$} 

Some properties of the $K_{1}$-$K_{2}$ intelligent states were studied 
in Refs.\ \cite{BHY_YH,GeGr}. We proceed here by using our general
analytic approach.
Here, $\vec{\beta} = (\eta,- {\rm i} ,0)$ and $B = \sqrt{1-\eta^{2}}$.
For $\eta^{2} \neq 1$, the corresponding analytic function is given 
by equation (\ref{an.2.6}) with 
        \begin{equation}
\tau_{\pm} = \mp \sqrt{ \frac{ 1-\eta }{ 1+\eta} } \, . 
\label{p.3.3.1}
        \end{equation}
For any $\eta > 0$ there exists $\epsilon > 0$ such that $|\tau_{\pm}| 
< (1+\epsilon)^{-1}$. Due to the features of the generation scheme,
we obtained in section \ref{sec:gener} the $K_{1}$-$K_{2}$ intelligent
states with $0 < \eta < 1$ and $\lambda = (k+l)B$. In principle,
normalizable $K_{1}$-$K_{2}$ intelligent states exist for any
$\eta > 0$ and for any complex eigenvalue $\lambda$. For $\eta = 1$,
one obtains the eigenstates of the lowering generator $K_{-}$. (These
states are known as the Barut-Girardello states \cite{BG71}; in the
context of the two-photon realization they are called even and odd 
coherent states \cite{DMM74}). 
For $l=0$, we obtain the SU(1,1) generalized coherent states 
$|\zeta_{0},k\rangle$ with $\zeta_{0} = -\tau_{+} =\tau_{-}$.

The parameters used in the general expressions are $\kappa = 2\tau_{+}$,
$t=4(1-\eta)/(1+\eta)$, $x=0$, $S_{+} = S_{-} = 2\eta/(1+\eta)$,
$Y=-4\eta/(1+\eta)^{2}$, $Z=8\eta^{2}(1-\eta)/(1+\eta)^{3}$. Then 
we obtain
        \begin{equation}
\langle K_{3} \rangle = \frac{1}{\eta} \left[ k +  
\frac{ l (l+2k) }{ 2k } \frac{ (1-\eta^2) }{ \eta^2 }\, \Theta \right]   
\label{p.3.3.2}
        \end{equation}
        \begin{eqnarray}
(\Delta K_{3})^{2} & = & [2l(l+2k)+k] \frac{ (1-\eta^2) }{ 2\eta^2 }
+ \frac{ l (l+2k) }{ 2k } \frac{ (1-\eta^2) (1-4k+\eta^2) }{ 2\eta^4 }\, 
\Theta \nonumber \\ & & - \left[ \frac{ l (l+2k) }{ 2k } 
\frac{ (1-\eta^2) }{ \eta^3 }\, \Theta \right]^2
\label{p.3.3.3}
        \end{eqnarray}
        \begin{equation}
\Theta = \frac{2k}{l} \left[ P_{l}^{(2k-1,0)}\left( 
\frac{ 2-\eta^2 }{ \eta^2 } \right) \right]^{-1} 
P_{l-1}^{(2k,1)}\left( \frac{ 2-\eta^2 }{ \eta^2 } \right) 
\;\;\;\;\;\;\; l \neq 0     \label{p.3.3.4} 
        \end{equation}
and $\Theta = 0$ for $l=0$.
The intensity correlation function $g^{(2)}$ is shown in figure~4 
versus $\eta$ for different values of $n$.
In the limit $\eta \ll 1$, we find
        \begin{equation}
\langle K_{3} \rangle \approx \frac{2n+1}{4\eta} \hspace{0.8cm} 
(\Delta K_{3})^{2} \approx \frac{2n+1}{8\eta^{2}} \hspace{0.8cm}
g^{(2)} \approx \frac{2n+3}{2n+1} .
\label{p.3.3.5}
        \end{equation}
We see that in this limit photon statistics is always super-Poissonian.
In the limit $\eta \rightarrow 1$, we define $\delta = 1-\eta^{2}$
and obtain for $\delta \ll 1$,
        \begin{equation}
\langle K_{3} \rangle \approx k + \frac{ (k+l)^{2} }{ 2k } \delta
\hspace{0.8cm}  (\Delta K_{3})^{2} \approx 
\frac{ (k+l)^{2} }{ 2k } \delta        \label{p.3.3.6}
        \end{equation}
        \begin{equation}
g^{(2)} \approx \left[ \left(n+\mbox{$\frac{1}{2}$} \right)^2 \delta 
\right]^{-1} \;\; {\rm for}\; k=1/4 \;\;\;\;\; {\rm and}\;\;\;\;\; 
g^{(2)} \approx \left(n+\mbox{$\frac{1}{2}$} \right)^2 \delta  \;\; 
{\rm for}\; k=3/4 .   \label{p.3.3.7}
        \end{equation}
\begin{figure}[htbp]\vspace*{-3.0cm}
\begin{minipage}[c]{.58\linewidth}
\epsfxsize=0.9\textwidth
\centerline{\epsffile{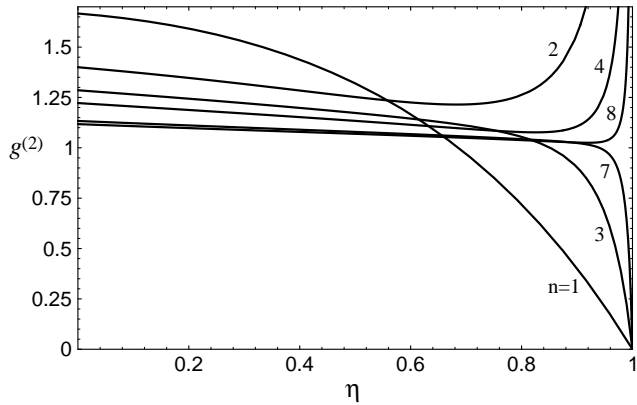}}\vspace*{-2.5cm}
\end{minipage}
\begin{minipage}[c]{.36\linewidth}
\caption{The intensity correlation function $g^{(2)}$ for 
the $K_{1}$-$K_{2}$ intelligent states versus $\eta$, for 
different values of $n$.}
\label{fig:4}
\end{minipage}
\end{figure}
We see here an additional example of the interesting phenomenon in 
the behaviour of photon statistics: the smaller is $\delta$, the 
stronger is antibunching for odd states and the stronger is bunching 
for even states. This behaviour is explained by the fact that in 
the limit $\eta \rightarrow 1$, the eigenstates of 
$\eta K_{1} - {\rm i} K_{2}$ reduce to the even and odd coherent states.

By the definition of the eigenstates,
$\langle \eta K_{1} - {\rm i} K_{2} \rangle = (k+l) \sqrt{1-\eta^{2}}$.
Therefore, we obtain
        \begin{equation}
\langle K_{1} \rangle = \frac{1}{\eta} (k+l) \sqrt{1-\eta^{2}}
\hspace{0.8cm}  \langle K_{2} \rangle = 0 .
\label{p.3.3.9}
        \end{equation}
The uncertainties of the field quadratures are
        \begin{equation}
(\Delta Q_{\pm})^{2} = \frac{2}{\eta} \left[ k +  
\frac{ l (l+2k) }{ 2k } \frac{ (1-\eta^2) }{ \eta^2 }\, \Theta 
\pm (k+l) \sqrt{1-\eta^{2}} \right]   .
\label{p.3.3.10}
        \end{equation}
The condition $(\Delta p)^{2} < (\Delta q)^{2}$ always holds here,
and squeezing can be observed only in the $p$ quadrature.
The quadrature uncertainty $(\Delta p)^{2}$ is shown in figure~5
versus $\eta$, for different values of $n$. 
In the limit $\eta \ll 1$, the $p$ quadrature is highly squeezed:
$(\Delta p)^{2} \rightarrow 0$ as $O(\eta)$, while the $q$ quadrature 
is very noisy: $(\Delta q)^{2} \approx (2n+1)/\eta$.
In the limit $\eta \rightarrow 1$ (i.e. $\delta \ll 1$), we obtain
        \begin{equation}
(\Delta Q_{\pm})^{2} \approx 2k \pm 2(k+l)\sqrt{\delta} + 
\frac{(k+l)^{2}}{k} \delta  .
\label{p.3.3.11}
        \end{equation}
\begin{figure}[htbp]\vspace*{-3.0cm}
\begin{minipage}[c]{.58\linewidth}
\epsfxsize=0.9\textwidth
\centerline{\epsffile{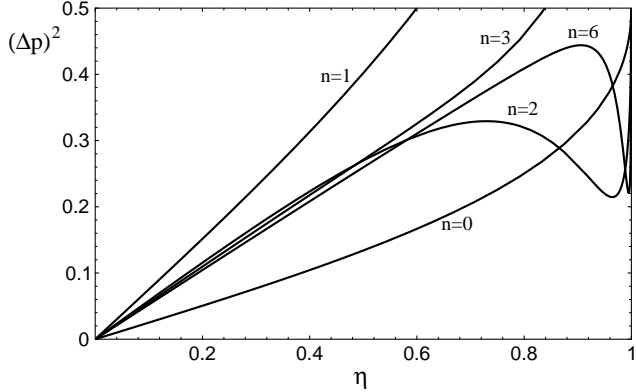}}\vspace*{-2.5cm}
\end{minipage}
\begin{minipage}[c]{.36\linewidth}
\caption{The quadrature uncertainty $(\Delta p)^{2}$ for 
the $K_{1}$-$K_{2}$ intelligent states versus $\eta$, for 
different values of $n$.}
\label{fig:5}
\end{minipage}
\end{figure}
For $k=1/4$ and $n \geq 2$, $(\Delta p)^{2}$ has a minimum 
($\approx 0.25$) for $\delta \approx 1/(2n+1)^{2}$. 
As $n$ increases, this minimum becomes sharper.  
We find that even states ($k=1/4$) are 
squeezed in the whole region $0 < \eta < 1$, while odd states 
($k=3/4$) are squeezed only for sufficiently small values of 
$\eta$.

According to the properties (\ref{intprop2}) of the intelligent
states, we have
        \begin{equation}
(\Delta K_{1})^{2} = \frac{ \langle K_{3} \rangle }{ 2 \eta }
\hspace{0.8cm} 
(\Delta K_{2})^{2} = \frac{ \eta \langle K_{3} \rangle }{ 2 } .
\label{p.3.3.12}
        \end{equation}
For $0 < \eta < 1$, the generator $K_{2}$ is always relatively
squeezed. However, it can be easily verified that absolute
quadratic squeezing is impossible for the $K_{1}$-$K_{2}$
ordinary intelligent states.

\subsubsection{Eigenstates of $\eta K_{2} + {\rm i} K_{3}$} 

Here, $\vec{\beta} = (0, \eta ,{\rm i})$ and 
$B = {\rm i} \sqrt{\eta^{2}+1}$. The corresponding analytic function 
is given by equation (\ref{an.2.6}) with
        \begin{equation}
\tau_{\pm} = \frac{ - \eta }{ 1 \pm \sqrt{\eta^{2}+1} } \, .
\label{an.2.3.2}
        \end{equation}
Note that $|\tau_{+}\tau_{-}| = 1$. For any $|\eta| < \infty$ there 
exists $\epsilon > 0$ such that $|\tau_{+}| < (1+\epsilon)^{-1}$.
Then $|\tau_{-}| > (1+\epsilon)^{-1}$ and the analyticity condition
requires that the spectrum is discrete, $\lambda = (k+l)B$. 
Equation (\ref{ist_yz}) shows that these eigenvalues naturally appear
in the generation scheme for the $K_{2}$-$K_{3}$ intelligent states.
For $\eta = 0$, these states reduce to the Fock states $|n\rangle$.
For $l=0$, we obtain the SU(1,1) generalized coherent states 
$|\zeta_{0},k\rangle$ with $\zeta_{0} = -\tau_{+}$.

Since $|\tau_{+}\tau_{-}| = 1$, we can use simple expressions
(\ref{p.2.17}), (\ref{p.2.18}) for the moments of the generator 
$K_{3}$. The parameters used in these expressions are $r=k+l$,
$h = (1+\sqrt{\eta^{2}+1})^2/\eta^{2}$, 
$t=4(\eta^{2}+1)/\eta^{2}$, $S_{+}S_{-} = -4/\eta^{2}$.
Then we obtain
        \begin{equation}
\langle K_{3} \rangle = (k+l) \sqrt{\eta^{2}+1}
\label{p.3.4.1}
        \end{equation}
        \begin{equation}
(\Delta K_{3})^{2} = \frac{k \eta^{2}}{2} \left[ 1 + \frac{l+2k}{k}
(\eta^{2}+1) \frac{ P_{l-1}^{(1,2k)}(2\eta^{2}+1) }{
P_{l}^{(0,2k-1)}(2\eta^{2}+1) } \right] \;\;\;\;\;\;\; l \neq 0 
\label{p.3.4.2}
        \end{equation}
and $(\Delta K_{3})^{2} = k \eta^{2}/2$ for $l=0$. In deriving
equation (\ref{p.3.4.2}) from (\ref{p.2.18}), we used the relation 
$P_{n}^{(\alpha,\beta)}(-x) = (-1)^{n} P_{n}^{(\beta,\alpha)}(x)$.
Note also that, by the definition of the eigenstates,
$\langle \eta K_{2} + {\rm i} K_{3} \rangle = 
{\rm i} (k+l) \sqrt{\eta^{2}+1}$.
This equation gives $\langle K_{2} \rangle = 0$ and the expession 
(\ref{p.3.4.1}) for $\langle K_{3} \rangle$.
In the limit $\eta \ll 1$, we find
        \begin{equation}
\langle K_{3} \rangle \approx \frac{2n+1}{4}
\left(1+\frac{\eta^{2}}{2}\right) \hspace{0.8cm}
(\Delta K_{3})^{2} \approx \frac{n^{2}+n+1}{8} \eta^{2} 
\label{p.3.4.4}
        \end{equation}
        \begin{equation}
g^{(2)} \approx 1-\frac{1}{n} + \frac{2n^{2}+4n+3}{4n^{2}} \eta^2 
\hspace{0.8cm} n \neq 0 .
\label{p.3.4.5}
        \end{equation}
Photon statistics is sub-Poissonian in accordance with the fact
that for $\eta \rightarrow 0$ the $K_{2}$-$K_{3}$ intelligent states
reduce to the Fock states $|n\rangle$. In the limit $\eta \gg 1$,
we obtain
        \begin{equation}
\langle K_{3} \rangle \approx \frac{2n+1}{4} \eta \hspace{0.8cm}
(\Delta K_{3})^{2} \approx \frac{2n+1}{8} \eta^{2} 
\label{p.3.4.6}
        \end{equation}
        \begin{equation}
g^{(2)} \approx \frac{2n+3}{2n+1} - \frac{2(2n-1)}{(2n+1)^{2}} 
\frac{1}{\eta} .
\label{p.3.4.7}
        \end{equation}
In this limit photon statistics is super-Poissonian.

According to the properties (\ref{intprop2}) of the intelligent
states, we have
        \begin{equation}
\langle K_{1} \rangle = \frac{2}{\eta} (\Delta K_{3})^{2}
\hspace{0.8cm}
(\Delta K_{2})^{2} = \frac{1}{\eta^{2}} (\Delta K_{3})^{2} .
\label{p.3.4.8}
        \end{equation}
Using equation (\ref{casimir}) for the Casimir operator, one finds
        \begin{equation}
\langle K_{1}^{2} \rangle = \langle K_{3}^{2} \rangle - 
\langle K_{2}^{2} \rangle - k(k-1) .    \label{p.3.4.9}
        \end{equation}
Then we obtain 
        \begin{equation}
(\Delta K_{1})^{2} = \frac{\eta^{2}-1}{\eta^{2}} (\Delta K_{3})^{2}
- \frac{4}{\eta^{2}} (\Delta K_{3})^{4} + (k+l)^{2} (\eta^{2}+1)
+ \frac{3}{16} .       \label{p.3.4.10}
        \end{equation}
The uncertainties of the field quadratures are
        \begin{equation}
(\Delta Q_{\pm})^{2} = 2 (k+l) \sqrt{\eta^{2}+1} \pm 
\frac{4}{\eta} (\Delta K_{3})^{2} .
\label{p.3.4.11}
        \end{equation}
Since $\eta$ is positive, the condition $(\Delta p)^{2} < 
(\Delta q)^{2}$ always holds here, and squeezing can be observed 
only in the $p$ quadrature. In the limit $\eta \ll 1$, we have
        \begin{equation}
(\Delta Q_{\pm})^{2} \approx \frac{2n+1}{2} \pm \frac{n^{2}+n+1}{2} 
\eta + \frac{2n+1}{4} \eta^{2} .
\label{p.3.4.12}
        \end{equation}
In this limit the $p$ quadrature is slightly squeezed only for
$n=0$. In the limit $\eta \gg 1$ the $p$ quadrature is strongly 
squeezed: $(\Delta p)^{2} \rightarrow 0$ as $O(1/\eta)$; while the 
$q$ quadrature is very noisy: $(\Delta q)^{2} \approx (2n+1)\eta$.
Numerical studies show that quadratic squeezing is possible only
for the generator $K_{2}$. In the limit $\eta \ll 1$, we have
        \begin{equation}
(\Delta K_{1})^{2} \approx (\Delta K_{2})^{2} 
\approx \frac{n^{2}+n+1}{8} .
\label{p.3.4.13}
        \end{equation}
There is no quadratic squeezing in this limit, except for weak
relative squeezing in $K_{2}$ for $n=0$. In the limit $\eta \gg 1$, 
the generator $K_{1}$ is very noisy, while $K_{2}$ is strongly
relatively squeezed:
        \begin{equation}
(\Delta K_{1})^{2} \approx \frac{2n+1}{8} \eta^{2} \hspace{0.8cm}
(\Delta K_{2})^{2} \approx \frac{2n+1}{8} .
\label{p.3.4.14}
        \end{equation}
Absolute quadratic squeezing is impossible for the $K_{2}$-$K_{3}$ 
ordinary intelligent states.

\section{Conclusions}

We have presented a scheme for the generation of the SU(1,1) 
intelligent states. This scheme employs quantum correlations created 
in a non-degenerate parametric amplifier between the vacuum and the 
squeezed vacuum. These quantum correlations (the entanglement) 
between the two light modes enable us to manipulate the state of one 
of the modes by a measurement of the photon number in the other. 
A powerful analytic method has been used for obtaining exact closed
expressions for quantum statistical properties of the intelligent
states. We have seen that these states can exhibit interesting 
nonclassical properties of strong antibunching and squeezing.
We have found that even states have a tendency to be squeezed, 
while odd states are more likely to be antibunched.

\section*{Acknowledgments}

CB gratefully acknowledges the financial help from the Technion
and thanks the Gutwirth family for the Miriam and Aaron Gutwirth
Memorial Fellowship.
AM was supported by the Fund for Promotion of Research
at the Technion, by the Technion VPR Fund --- R. and M. Rochlin
Research Fund, and by GIF --- German-Israeli Foundation for 
Research and Development.

\newcommand{\bisk}{\vspace*{-2mm}}
\begin{flushleft}

\end{flushleft}

\end{document}